\documentclass[final,3p,times,2024]{elsarticle}

\usepackage[T1]{fontenc}
\usepackage[utf8]{inputenc}

\usepackage{enumitem}
\usepackage{subfiles}       
\usepackage{booktabs}       
\usepackage{amsmath,xfrac}
\usepackage{multirow}
\usepackage{float}
\usepackage{dblfloatfix}    
\usepackage{bookmark}       
\usepackage{lineno}         
\usepackage{siunitx}        
\usepackage{graphicx}
\graphicspath{{figures/}}
\usepackage{placeins}
\usepackage{float}
\usepackage{multirow}
\usepackage{tabularx}

\usepackage{nth}
\usepackage{url}
\usepackage{longtable}     
\usepackage{booktabs}
\usepackage{multirow}
\usepackage{xcolor}    
\usepackage[flushleft]{threeparttable} 
\usepackage{caption,nameref}
\usepackage{subcaption}
\usepackage[version=4,arrows=pgf-filled,
textfontname=sffamily,
mathfontname=mathsf]{mhchem}
\usepackage{diagbox,makecell}
\usepackage{siunitx}
\usepackage{multirow,diagbox}
\usepackage{makecell}

\setcitestyle{square}
\usepackage{dcolumn,makecell}
\usepackage[utf8]{inputenc}
\usepackage{tikz}
\usetikzlibrary{arrows.meta, positioning, fit, backgrounds}
\usepackage{caption}
\usepackage{algorithm}
\usepackage{algpseudocode}
\algrenewcommand\algorithmicrequire{\textbf{Input:}}
\algrenewcommand\algorithmicensure{\textbf{Output:}}

\algdef{SE}[FORLOOP]{ForLoop}{EndForLoop}[1]%
{\algorithmicfor\ #1\ \algorithmicdo}%
{\algorithmicend\ \algorithmicfor}

\usepackage{bm, comment} 
\usepackage{hyperref}
\hypersetup{
    colorlinks=true,
    linkcolor=blue,
    filecolor=magenta,      
    urlcolor=cyan,
    pdftitle={Overleaf Example},
    pdfpagemode=FullScreen,
    }

\biboptions{sort&compress} 
\usepackage{setspace}
\usepackage{multirow}

\graphicspath{{./Figures/}}

\begin{document}

\sisetup{
    inter-unit-product = ,
    per-mode=symbol
    }

\begin{frontmatter}

\title{Timescale-aware surrogate-assisted multi-objective optimization of battery cell design for energy density, fast charging, and degradation}

\author[1]{Qingbo~Zhu\corref{cor1}}
\ead{qingbo.zhu@chalmers.se}
\author[1]{Changfu~Zou}
\ead{changfu.zou@chalmers.se}
\author[2]{Yicun~Huang}
\ead{yicun.huang@it.uu.se}
\author[1]{Chunqiu~Xia}
\ead{chunqiu@chalmers.se}
\author[1]{Torsten~Wik\corref{cor1}}
\ead{tw@chalmers.se}

\address[1]{Department of Electrical Engineering, Chalmers University of Technology, Gothenburg, 412 96, Sweden.}
\address[2]{Department of Information Technology, Uppsala University, Uppsala, 751 05, Sweden.}
\cortext[cor1]{Corresponding authors. Address: H\"{o}rsalsv\"{a}gen 11, SE-412 96 Gothenburg, Sweden.} 

\begin{abstract}
Battery cell design must balance energy density, fast charging, and degradation, yet these metrics evolve over different time scales and are costly to optimize jointly. We develop a timescale-aware surrogate-assisted framework that evaluates beginning-of-life volumetric energy density and 10--80\% charging time together with state-of-health (SOH) loss over 200 cycles. Physics-based simulations of 1501 cell designs generate 1427 quality-controlled samples across 12 manufacturing-relevant parameters. Objective-specific surrogates support total-order Sobol analysis, which reveals distinct parameter rankings across the three metrics. A cross-objective rank-union strategy then retains variables influential to at least one objective before evolutionary Pareto optimization. Re-evaluation of the optimized candidates using the original physics-based models confirms designs that outperform the reference cell in all three metrics. Among these jointly improving candidates, the objective-wise best solutions, attained by different designs, can reduce SOH loss by 99.31\%, increase volumetric energy density by 12.49\%, and shorten charging time by 28.73\%. The framework makes Pareto exploration across disparate electrochemical timescales computationally tractable, thereby enabling systematic multi-objective optimization of battery cell design.
\end{abstract}

\end{frontmatter}


\section{Introduction}
\label{sec:introduction}
Electrification of road transport is central to reducing petroleum dependence and transport-sector emissions. Its scale is already substantial: global electric-vehicle (EV) battery deployment reached approximately 1.2\,TWh in 2025, nearly 30\% above the 2024 level~\cite{IEA2026GlobalEVOutlook}. Lithium-ion batteries remain the dominant energy-storage technology for this transition because they combine high energy density, high efficiency, and comparatively long service life~\cite{deng2020requirements,zhang2021comparative}. 
Continued growth, however, places increasingly stringent demands on the cell itself. More energy must be stored within a constrained vehicle volume, charging must approach the time scale of conventional refueling, and degradation must remain sufficiently slow to avoid premature replacement and the associated material, economic, and environmental burdens.

These requirements are physically coupled. Increasing electrode thickness or reducing porosity can raise active-material loading and volumetric energy density, but the former lengthens and the latter constricts ionic-transport pathways~\cite{parikh2020correlating,boyce2022exploring}. The resulting concentration gradients and polarization become especially restrictive during fast charging. Particle-size reduction can improve solid-state diffusion and rate capability, yet the accompanying increase in active surface area also enlarges the electrode--electrolyte interface available for parasitic reactions~\cite{gao2017parasitic,jain2022nanostructuring}. Such effects accumulate over cycling: a design that performs well during a beginning-of-life (BoL) discharge or charging event may not retain that advantage after repeated operation. Physics-based aging analyses have consequently identified a persistent trilemma among energy density, fast charging, and cycle life~\cite{yang2018trilemma}. Cell design must therefore be evaluated through application-level performance metrics, not through any single geometric or material descriptor.

Early model-based design studies concentrated primarily on energy storage and rate capability. De et al.~\cite{de2013model} optimized multiple electrode and material parameters to maximize energy density, while Kim et al.~\cite{kim2020optimization} combined design of experiments with response-surface optimization for specific energy density. More recent manufacturing-oriented studies have used physics-based simulations and machine-learning surrogates to optimize electrode density, tortuosity, electronic conductivity, active surface area, or energy--power trade-offs~\cite{Duquesnoy2023multiobjective,duquesnoy2024highperformance}. These formulations provide valuable guidance for electrode manufacturing and microstructure design. Their objectives, however, are predominantly intermediate descriptors or cell responses evaluated over one or a few operating events.

Degradation-aware cell design has also been investigated, including through explicit multi-cycle simulation. A multiphysics capacity-fade model was combined with non-dominated sorting genetic algorithm II (NSGA-II) to optimize discharge specific energy, specific power, and residual capacity, and the optimized designs were further examined over 750 cycles~\cite{liu2017optimal}. Lin and Lu~\cite{lin2018framework} minimized capacity fade after 100 cycles subject to energy- and power-density requirements, with the associated cyclable-lithium loss analyzed over the first 200 cycles. Together with the energy--charging--life analysis of Yang and Wang~\cite{yang2018trilemma}, these studies establish an important point: aging is not an operational correction that can simply be appended after a BoL design has been selected. Electrode dimensions and particle properties that favor energy or power can shift once degradation is included.

Surrogate-assisted methods have subsequently reduced the cost of combining physics-based models with iterative design. Cui et al.~\cite{cui2022codesign} used adaptive surrogate modeling to jointly optimize electrode design and charging control, minimizing charging time at prescribed health requirements. Ma et al.~\cite{ma2024multiobjective} developed a P2D-model-based surrogate and NSGA-II framework targeting energy density, fast charging, high-rate discharge, and lifespan. Their Pareto construction primarily used low-rate discharge energy density, high-rate discharge energy density, and the energy charged within a fixed 10-min interval; lifespan was examined through lithium-plating and aging analyses rather than represented by a separately learned, cycle-accumulated state-of-health (SOH) objective. Ju et al.~\cite{ju2026multiobjective} coupled an electrochemical--thermal--mechanical--side-reaction aging model with surrogate modeling, SHAP analysis, and evolutionary optimization to balance energy density against capacity loss, but did not include charging time as an optimization objective.

What remains unresolved is not whether degradation matters, but how to carry an explicitly cycle-accumulated degradation response into a broad design search together with energy storage and charging speed. The three quantities are aligned through the same cell-design vector, yet their physics-based evaluations are intrinsically separated in time scale. Energy density and charging time can be determined from BoL operating simulations; SOH loss requires repeated cycling. Embedding the full aging calculation in every sensitivity-analysis and optimization evaluation is expensive, whereas representing lifetime only through a BoL proxy can obscure design-dependent degradation. A further difficulty arises during dimension reduction. Because the influential variables need not be the same for all objectives, screening parameters against a single response can remove variables that are important to another competing objective. The consequences of such reduction for attainable Pareto performance and the stability of the resulting design recommendations have received comparatively limited attention.

This study develops a timescale-aware framework for optimizing battery cell design with respect to stack-level volumetric energy density, 10--80\% charging time, and SOH loss accumulated over 200 cycles. Twelve manufacturing-relevant parameters define a common design space around an experimentally parameterized graphite--SiO$_x$/NMC811 reference configuration. Volumetric energy density and charging time are evaluated using BoL Doyle--Fuller--Newman (DFN) simulations, whereas the degradation response is obtained from 200-cycle simulations using a single-particle model with electrolyte dynamics (SPMe) and explicit degradation mechanisms. After numerical quality control, the three simulation workflows yield 1427 aligned candidate designs.
The principal contributions are as follows:
\begin{itemize}
\item A timescale-aware, objective-specific surrogate-modeling strategy is developed. The three targets share the same 12-dimensional input space but retain their respective physics-based simulation protocols, target distributions, and independently selected surrogate models.
\item Total-order Sobol indices are evaluated separately for SOH loss, volumetric energy density, and charging time. A rank-union strategy then retains variables that are influential for any objective, avoiding dimension reduction dominated by a single performance metric.

\item NSGA-II is applied across a sequence of nested, sensitivity-informed design spaces. This construction quantifies how the number of adjustable parameters affects the attainable three-objective Pareto performance. 

\item Pareto candidates from every reduced design space are returned to the original high-fidelity simulations. This physics-based re-evaluation identifies feasible designs that simultaneously outperform the reference configuration and reveals performance saturation and parameter stability as the optimization space expands.
\end{itemize}

\begin{figure}[!htbp]
    \centering
    \includegraphics[width=\linewidth]{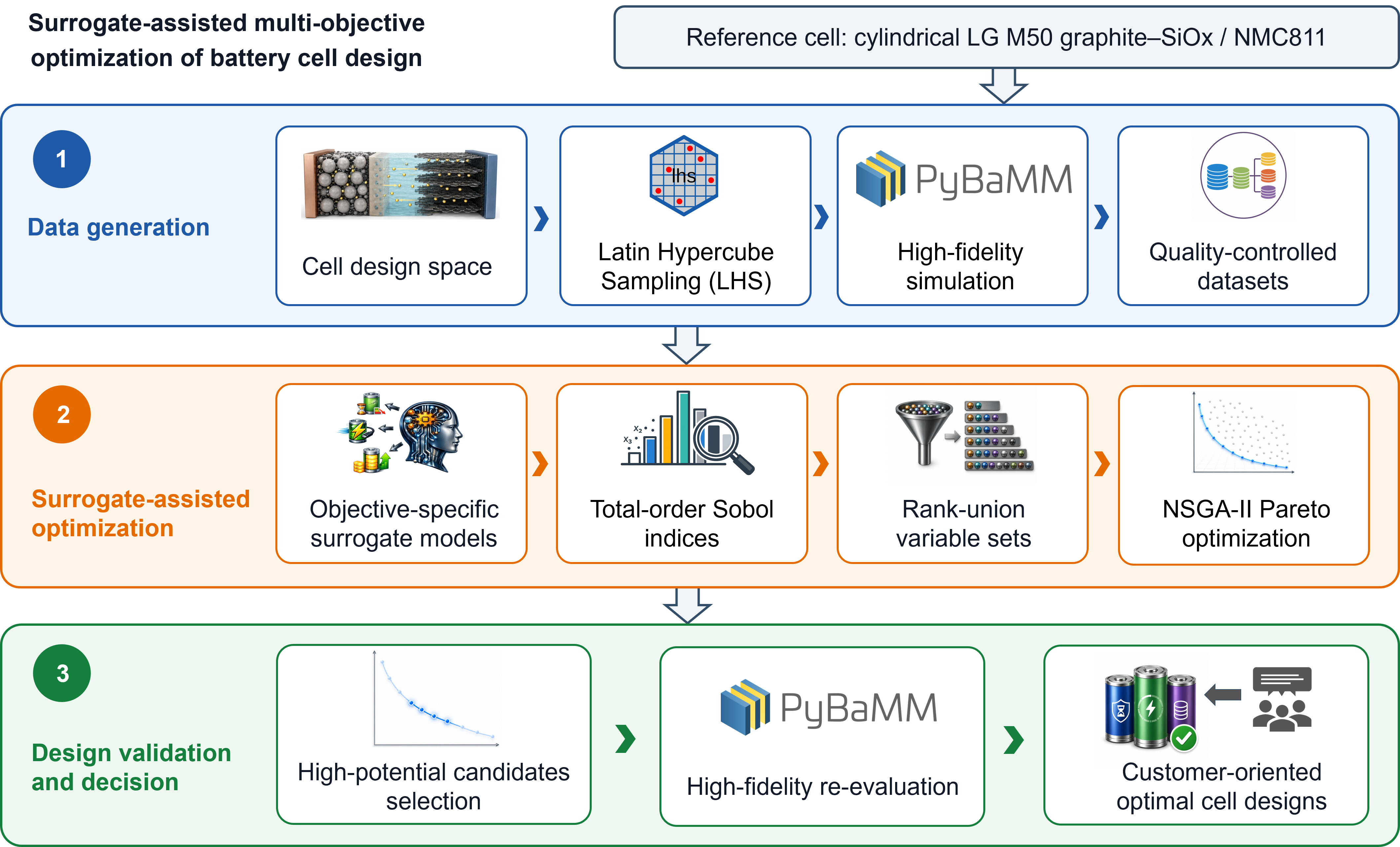}
    \caption{Workflow of the surrogate-assisted multi-objective optimization and PyBaMM confirmation procedure.}
    \label{fig:workflow}
\end{figure}

\section{Performance metrics and design space}
This section defines the cell-design problem, including the performance metrics used for evaluation and the parameter space over which candidate designs are generated.

\subsection{Performance metrics}\label{sec:performance_metrics}
Cell design is governed by multiple and potentially competing requirements. A design that increases electrode loading may improve energy density but also lengthen transport pathways, thereby impairing fast-charging capability and accelerating degradation. We therefore consider three metrics that jointly capture the principal cell-level requirements relevant to electric-vehicle (EV) applications: degradation rate, energy density, and fast-charging time.

{\bf Degradation rate.} 
SOH degradation is a key cell-design metric because its progression determines the attainable service life. A longer service life reduces battery replacement frequency and life-cycle cost, while lowering the material and energy demands associated with battery production and end-of-life treatment, thereby improving both cost efficiency and sustainability.

For battery cells in EVs, end of life is commonly defined as the time or number of cycles at which the cell state of health (SOH) decreases to 75\%--80\% of its initial value. Lifetime should therefore, in principle, be considered directly during cell design. Battery degradation, however, evolves considerably more slowly than the intercalation and diffusion processes within a cell. Experimentally aging a cell to end of life can require several years and substantial electrical energy. High-fidelity battery simulations can significantly reduce this experimental burden by capturing the coupled effects of cell design and long-term operation, but multiphysics simulations over thousands of cycles remain computationally prohibitive when a large number of candidate designs must be evaluated. 

For computational tractability, and to demonstrate the proposed optimization framework, we evaluate the accumulated SOH loss over the first 200 cycles rather than simulating every candidate design to end of life. The SOH degradation rate over the first 200 cycles is defined as
\begin{equation}
r_{\mathrm{SOH}}
=
\frac{Q_{\mathrm{init}}-Q_{\mathrm{final}}}
{Q_{\mathrm{init}}}.
\label{eq:soh_200}
\end{equation}
where \(Q_{\mathrm{init}}\) and \(Q_{\mathrm{final}}\) are the cell capacities calibrated at the beginning of life (BoL) and after 200 charge--discharge cycles, respectively. The selected horizon provides a consistent basis for comparing design-dependent degradation and can be replaced by a longer aging horizon or another lifetime criterion without altering the optimization framework. A lower value of \(r_\text{SOH}\) indicates better resistance to degradation and is therefore preferred.

\textbf{Energy density.}
Volumetric energy density quantifies the deliverable energy per unit volume of a battery cell. It is particularly important for EV applications because the available battery-pack volume is constrained by vehicle architecture; accommodating more energy within a given volume directly supports longer driving range and greater flexibility in pack integration. We therefore adopt volumetric energy density as the energy-storage performance metric, which is defined as
\begin{equation}
e_V
= 
\frac{1}{3600}
\frac{\int_{0}^{t_\mathrm{f}}
|I(\tau)|V(\tau)\,d\tau }
{v_{\mathrm{cell}}
} \,,
\label{eq:e_init}
\end{equation}
where \(V\), \(I\), and $v_{\mathrm{cell}}$ denote the terminal voltage, applied current, and cell volume, respectively. \(t_\mathrm{f}\) is the operating duration corresponding to a full charge or discharge at the BoL. The factor \(1/3600\) converts the integrated energy from watt-seconds to watt-hours. 

In~\eqref{eq:e_init}, the cell volume is approximated by the stack volume of the electrode sandwich and current collectors
\begin{equation}
v_{\mathrm{cell}} \approx
1000 A\left( L_{\mathrm{pcc}} + L_{\mathrm{p}}
+ L_{\mathrm{s}} + L_{\mathrm{n}} + L_{\mathrm{ncc}}
\right),
\label{eq:v_cell_approx}
\end{equation} 
where \(A\) is the equivalent electrode area; \(L_{\mathrm{p}}\), \(L_{\mathrm{s}}\), and \(L_{\mathrm{n}}\) are the positive-electrode, separator, and negative-electrode thicknesses; and \(L_{\mathrm{pcc}}\) and \(L_{\mathrm{ncc}}\) are the positive and negative current-collector thicknesses, respectively. The factor \(1000\) converts the volume from m\(^3\) to L, such that \(e_V\) is expressed in Wh\,L\(^{-1}\). This approximation excludes the casing, tabs, external interconnects, and other packaging components, and is therefore used for consistent relative comparison among candidate cell designs rather than as an absolute packaged-cell energy density. 

\textbf{Fast-charging time.}
Fast-charging capability determines how rapidly a substantial fraction of the usable cell capacity can be restored. It is therefore important for reducing charging downtime and improving the practical usability of EVs. In this work, fast-charging performance is quantified by the time required to increase the cell SOC from 10\% to 80\%, denoted by \(t_\textrm{chg}\). This SOC interval represents the practically relevant rapid-charging window while excluding the near-full-SOC region, where voltage constraints and current tapering increasingly dominate the charging process~\cite{wikner2018extending,zhang2025machine}.

Because candidate designs may have different capacities, the target charge throughput is defined relative to the fresh-cell capacity \(Q_{\mathrm{init}}\). Let \(t_{0}\) denote the time at which charging begins from 10\% SOC. The fast-charging time is defined as
\begin{equation}
t_{\mathrm{chg}}
=
\min_{\Delta t\geq 0}
\left\{
\Delta t:\,
\frac{1}{3600}
\int_{t_{0}}^{t_{0}+\Delta t}
|I_{\mathrm{chg}}(\tau)|\,\mathrm{d}\tau
\geq
0.7Q_{\mathrm{init}}
\right\},
\label{eq:t_chg}
\end{equation}
where \(I_{\mathrm{chg}}\) denotes the charging current. A lower value of \(t_{\mathrm{chg}}\) indicates stronger fast-charging capability and is therefore preferred. Candidate designs that do not reach the required charge throughput under the prescribed charging protocol are treated as infeasible.

\subsection{Design-space definition and sampling}
\label{sec:design_space_sampling}

To optimize cell design with respect to the three performance metrics defined in Section~\ref{sec:performance_metrics}, we select 12 manufacturing-relevant cell-design parameters associated with cell geometry, porous-electrode transport, active-material storage capacity, solid-state diffusion, and electrolyte transport. Related physics-based and machine-learning-assisted studies have shown that electrode geometry, porous-electrode structure, and material and transport properties should be considered jointly when designing cells for energy and rate performance~\cite{de2013model,Hui2021,Duquesnoy2023multiobjective}.

For each electrode, the selected parameters are electrode thickness, porosity, Bruggeman coefficient, maximum lithium concentration in the active material, and active-material particle radius. Electrode thickness and porosity characterize active-material loading, electrolyte-filled pore volume, and through-plane transport length; the Bruggeman coefficient captures porosity-dependent effective transport; the maximum lithium concentration represents active-material lithium-storage capacity; and particle radius sets the characteristic solid-state diffusion length scale, particularly under high-rate operation~\cite{Zhao2015,Tian2019,Lu2020}. The separator Bruggeman coefficient and initial electrolyte concentration extend the design description to separator transport and electrolyte-state effects. Collectively, these 12 parameters strongly influence active-material utilization, SOH degradation rate, volumetric energy density, and fast-charging time. The parameters and their physical relevance are summarized in Table~\ref{tab:design_parameters}.

\begin{table}[htbp]
\centering
\caption{Cell-design parameters defining the optimization space and their physical relevance.}
\label{tab:design_parameters}
\begin{tabular}{llll}
\toprule
Cell component & Symbol & Parameter & Physical relevance \\
\midrule
\multirow{5}{*}{Positive electrode}
& \(L_{\mathrm{p}}\) 
& Electrode thickness 
& Active-material loading and transport length \\

& \(\varepsilon_{\mathrm{p}}\) 
& Electrode porosity 
& Electrolyte volume fraction and ionic transport \\

& \(b_{\mathrm{p}}\) 
& Bruggeman coefficient 
& Porosity-dependent effective transport \\

& \(c_{\mathrm{p,max}}\) 
& Maximum lithium concentration 
& Active-material lithium-storage capacity \\

& \(R_{\mathrm{p}}\) 
& Particle radius 
& Solid-state diffusion length scale \\
\midrule

\multirow{5}{*}{Negative electrode}
& \(L_{\mathrm{n}}\) 
& Electrode thickness 
& Active-material loading and transport length \\

& \(\varepsilon_{\mathrm{n}}\) 
& Electrode porosity 
& Electrolyte volume fraction and ionic transport \\

& \(b_{\mathrm{n}}\) 
& Bruggeman coefficient 
& Porosity-dependent effective transport \\

& \(c_{\mathrm{n,max}}\) 
& Maximum lithium concentration 
& Active-material lithium-storage capacity \\

& \(R_{\mathrm{n}}\) 
& Particle radius 
& Solid-state diffusion length scale \\
\midrule

Separator
& \(b_{\mathrm{s}}\) 
& Bruggeman coefficient 
& Tortuosity-dependent separator transport \\

Electrolyte
& \(c_{\mathrm{e},0}\) 
& Initial electrolyte concentration 
& Electrolyte transport and concentration polarization \\
\bottomrule
\end{tabular}
\end{table}

The design space is constructed by allowing each of the 12 parameters to vary independently by up to 20\% above or below its reference value. Specifically, let \(x_{j,0}\) denote the reference value of the \(j\)-th design parameter. A candidate value \(x_j\) is expressed as
\begin{equation}
x_j
=
x_{j,0}\left(1+r_j\right),
\quad
r_j\in[-20\%,\,20\%],
\quad
j=1,\ldots,12,
\label{eq:relative_perturbation}
\end{equation}
where \(r_j\) is the relative perturbation from the reference design. The symmetric \(\pm20\%\) bounds permit substantial yet controlled variation around the reference cell while limiting parameter combinations that are physically implausible or prone to numerical failure. These bounds define the computational design space used for subsequent data generation, surrogate-model development, and optimization. They should not be interpreted as empirically derived manufacturing tolerances or uncertainty bounds.

Exhaustively discretizing the resulting 12-dimensional continuous space would require an impractically large number of parameter combinations. Latin hypercube sampling (LHS) is therefore used to generate a finite set of candidate designs. LHS divides the range of each parameter into \(N^{\prime}=1500\) intervals of equal probability and draws one value from each interval. The independently stratified values are then combined across the 12 dimensions to form \(N^{\prime}\) design vectors. Compared with simple random sampling, LHS provides more uniform marginal coverage for a fixed sample size and reduces the likelihood that portions of the prescribed parameter ranges are poorly represented~\cite{mckay2000comparison}.

A total of \(N^{\prime}=1500\) LHS designs are generated. This sample size represents a practical compromise between coverage of the 12-dimensional design space and the computational cost of the subsequent physics-based simulations. With the reference parameter set included, we obtain 1501 candidate designs in total.

Physical consistency is enforced when constructing the candidate parameter sets. For each porous electrode \(\lambda\in\{\mathrm{p},\mathrm{n}\}\), the volume fractions satisfy
\begin{equation}
\varepsilon_{\mathrm{am},\lambda}
+
\varepsilon_{\lambda}
+
\varepsilon_{\mathrm{inam},\lambda}
=
1,
\qquad
\lambda\in\{\mathrm{p},\mathrm{n}\},
\label{eq:electrode_volume_fraction}
\end{equation}
where \(\varepsilon_{\mathrm{am},\lambda}\) and \(\varepsilon_{\mathrm{inam},\lambda}\) denote volume fractions of the active and inactive solid materials, respectively. The inactive solid phase represents non-active electrode constituents, primarily the binder and conductive additive. Because the electrode porosity \(\varepsilon_\lambda\) is varied as a design parameter while \(\varepsilon_{\mathrm{inam},\lambda}\) is fixed at its reference value, the active-material fraction is updated according to
\begin{equation}
\varepsilon_{\mathrm{am},\lambda}
=
1-\varepsilon_{\lambda}-\varepsilon_{\mathrm{inam},\lambda}.
\label{eq:active_material_fraction}
\end{equation}
This construction ensures that every sampled electrode composition satisfies the required volume-fraction balance before physics-based simulation.

\section{Data generation}\label{sec:data_generation}
The 1501 candidate designs defined in Section~\ref{sec:design_space_sampling} are evaluated using electrochemical models implemented in PyBaMM, an open-source battery-modeling and simulation environment~\cite{sulzer2021python}. The reference values of the 12 selected design parameters (\(x_{j,0}\)), together with the parameters held fixed throughout the design space, are taken from an experimentally parameterized cylindrical LG M50 cell comprising a graphite--SiO$_x$ negative electrode and an NMC811 positive electrode~\cite{chen2020development}. These parameters define the reference configuration. Once the selected parameters are perturbed, the resulting parameter sets represent new cell designs. The principal geometric and operating quantities of the reference configuration are listed in Table~\ref{tab:reference_cell}.

The model fidelity is selected according to the physical processes and time scale associated with each performance metric. The DFN model is used for the BoL capacity, energy-density, and fast-charging evaluations, each of which is performed only once per candidate design. The substantially longer aging calculation is instead based on the SPMe, coupled with explicit degradation submodels. Throughout the simulations, the applied current corresponding to each stated C-rate is calculated using the fixed 5.0\,Ah capacity of the reference configuration and is therefore identical across all candidate designs.

\begin{table}[th]
\centering
\caption{Principal numerical quantities of the reference cell configuration.}
\label{tab:reference_cell}
\begin{tabular}{lcc}
\toprule
Quantity & Value & Unit \\
\midrule
Nominal capacity & 5.0 & Ah \\
Lower voltage cut-off & 2.5 & V \\
Upper voltage cut-off & 4.2 & V \\
Electrode height & 0.065 & m \\
Electrode width & 1.58 & m \\
Equivalent electrode area & 0.1027 & m$^2$ \\
Positive current-collector thickness & 16.3 & $\mu$m \\
Positive-electrode thickness & 75.6 & $\mu$m \\
Separator thickness & 12 & $\mu$m \\
Negative-electrode thickness & 85.2 & $\mu$m \\
Negative current-collector thickness & 11.7 & $\mu$m \\
Reference stack thickness & 200.8 & $\mu$m \\
Reference stack volume & 0.02062 & L \\
\bottomrule
\end{tabular}
\end{table}

\subsection{Initial capacity and energy density}\label{sec:energy_density_generation}
The initial capacity and volumetric energy density of each candidate design are evaluated at the BoL using the DFN model~\cite{doyle1993modeling,li2026fast}. The DFN formulation retains the through-thickness variations in electrolyte concentration and potential, solid-phase potential, and interfacial reaction current density that determine the terminal-voltage response. Its computational cost remains manageable here because the capacity-calibration sequence is performed only once for each design.

Following equilibrium initialization, the cell first undergoes a full constant-current--constant-voltage (CCCV) charge. It is charged at 0.5C until the terminal voltage reaches 4.2\,V and is subsequently held at 4.2\,V until the current decreases to C/100. After a 30\,min rest, the cell is discharged at 0.1C to the lower voltage cut-off of 2.5\,V. The current and voltage trajectories are recorded at 1\,Hz. The initial capacity is defined as the mean of the charge and discharge throughputs:
%
%
\begin{equation}
Q_{\mathrm{init}}
=
\frac{1}{2\times 3600}
\left(
\int_{0}^{T_{\mathrm{chg}}^{\mathrm{cal}}}
\left|I_{\mathrm{chg}}(\tau)\right|\,\mathrm{d}\tau
+
\int_{0}^{T_{\mathrm{dis}}^{\mathrm{cal}}}
\left|I_{\mathrm{dis}}(\tau)\right|\,\mathrm{d}\tau
\right),
\label{eq:initial_capacity}
\end{equation}
where \(T_{\mathrm{chg}}^{\mathrm{cal}}\) and \(T_{\mathrm{dis}}^{\mathrm{cal}}\) denote the durations of the calibration charge and discharge, respectively. Averaging the charge and discharge throughputs reduces the directional bias introduced by voltage-based termination in the presence of polarization and voltage hysteresis.

Using the same current and voltage trajectories, the BoL volumetric energy density is defined as the mean of the charge and discharge energy throughputs normalized by the stack volume:
\begin{equation}
e_V
=
\frac{1}{2\times 3600\,v_{\mathrm{cell}}}
\left(
\int_{0}^{T_{\mathrm{chg}}^{\mathrm{cal}}}
\left|I_{\mathrm{chg}}(\tau)\right|
V_{\mathrm{chg}}(\tau)\,\mathrm{d}\tau
+
\int_{0}^{T_{\mathrm{dis}}^{\mathrm{cal}}}
\left|I_{\mathrm{dis}}(\tau)\right|
V_{\mathrm{dis}}(\tau)\,\mathrm{d}\tau
\right),
\label{eq:e_init}
\end{equation}
where \(v_{\mathrm{cell}}\) is the stack volume defined in Eq.~\eqref{eq:v_cell_approx}. Because the casing, tabs, and other packaging components are excluded from \(v_{\mathrm{cell}}\), \(e_V\) is a stack-level metric used for consistent comparison among candidate designs rather than the energy density of a fully packaged commercial cell.

\subsection{Fast-charging response}\label{sec:fast_charging_generation}
Fast-charging performance is evaluated independently at the BoL, after the capacity and energy-density calibration. The DFN model is retained for this evaluation because fast charging produces pronounced through-thickness gradients in electrolyte concentration and phase potentials, together with a non-uniform distribution of interfacial reaction current density. These spatial variations directly affect when the upper-voltage constraint is encountered and are not fully resolved by reduced-order electrode models.

For each candidate design, the battery model is reinitialized at 10\% SOC and charged at a constant current of 2C. The charge is terminated when the cell voltage reaches the upper cut-off of 4.2\,V. The fast-charging time \(t_{\mathrm{chg}}\) is obtained from the simulated current trajectory according to Eq.~\eqref{eq:t_chg}. A candidate design is classified as infeasible under the prescribed fast-charging protocol if the voltage cut-off is reached before \(0.7Q_{\mathrm{init}}\) has been delivered.

\subsection{Aging simulation}\label{sec:aging_generation}
The aging calculation differs from the two BoL evaluations in both time scale and model configuration. Simulating 200 cycles for every candidate design using the DFN model would impose a disproportionate computational cost. The SPMe is therefore used as the electrochemical backbone~\cite{moura2016battery} and is coupled with degradation submodels adopted from O'Kane \emph{et al.}~\cite{okane2022lithium}. The SPMe itself does not describe battery aging; degradation arises from the additional submodels coupled to its electrochemical states. Three degradation mechanisms are included: solvent-diffusion-limited solid--electrolyte interphase (SEI) growth, stress-driven loss of active material, and the reduction of negative-electrode porosity caused by SEI accumulation. These mechanisms account for loss of cyclable lithium, mechanically induced active-material degradation, and degradation-induced changes in porous-electrode transport, respectively. 

Lithium plating is another major degradation mechanism that may occur in lithium-ion cells under real-world operating conditions, but it is not included in the present model. This choice reflects the operating premise that next-generation battery management systems (BMSs) may suppress lithium plating and extend battery lifetime by intelligently adjusting the battery usage profile~\cite{tian2021detecting,huang2022onboard,zhang2025machine}. 
Consequently, even when parameterized using the LG M50 reference values, the simulated reference design will generally exhibit a longer lifetime than a commercial LG M50 cell. The coupled model is therefore not intended to reproduce every degradation pathway present in a commercial cell. Instead, it provides a common mechanistic basis for comparing designs with different geometries, active-material properties, particle dimensions, and transport parameters. Accordingly, \(r_{\mathrm{SOH}}\) represents the relative degradation response under the specified model and cycling protocol and should not be interpreted as a universal prediction of commercial-cell lifetime.

\begin{figure}[htbp]
    \centering
    \includegraphics[width=\linewidth, trim={0 0 0 2mm},clip]{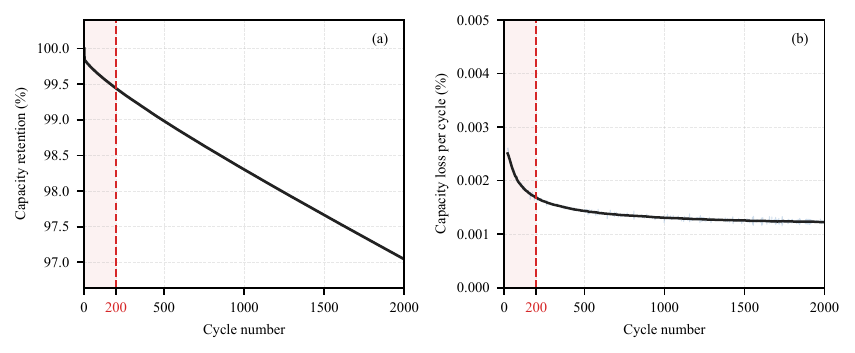}
    \caption{Aging response of the reference design under repeated 1C charge--discharge cycling. 
(a) Trajectory of the actual cell capacity  of the reference design. 
(b) Capacity loss per cycle, calculated as the difference in cell capacity between two consecutive cycles, normalized by \(Q_{\mathrm{init}}\).
The red shaded regions highlight the first 200 cycles selected as the common horizon for evaluating degradation across the candidate designs. 
}
\label{fig:baseline-aging-horizon}
\end{figure}

Figure~\ref{fig:baseline-aging-horizon} presents the capacity retention and cycle-to-cycle capacity loss of the reference design over 2000 cycles. Under the degradation mechanisms considered, the cycle-to-cycle capacity loss is largest during the early cycling period and generally decreases thereafter. The first 200 cycles therefore provide a common early-aging window in which the degradation responses of the candidate designs can be distinguished while keeping the evaluation of the full design space computationally tractable. This horizon is a modeling choice for evaluating the optimization framework, not an assumed end-of-life criterion. Data-driven models can efficiently predict battery lifetime from early cycling data~\cite{xiong2020lithium,cui2024ultra}, 
but their development requires end-of-life-labeled data. Such lifetime prediction is beyond the primary scope of this work and is therefore not pursued.

For each candidate design, the virtual cell undergoes CCCV charging at 1C until the terminal voltage reaches 4.2\,V, after which the voltage is held at 4.2\,V until the current decreases to 0.1C. The cell is then discharged at 1C to 2.5\,V. This cycling sequence is repeated for 200 cycles. The final capacity $Q_\mathrm{final}$ is subsequently evaluated using the same charge--discharge calibration protocol employed to determine \(Q_{\mathrm{init}}\), as described in Section~\ref{sec:energy_density_generation}. The two capacity measurements are then used to evaluate SOH degradation \(r_{\mathrm{SOH}}\).

\subsection{Dataset construction and numerical quality control}
\label{sec:dataset_construction}
For each candidate design \(i\), the outputs from the aging, energy-density, and fast-charging evaluations are combined through their common vector of design parameters \(\mathbf{x}_i\). A candidate design is retained only if the simulations required for its three performance evaluations converge and yield finite metric values, the required capacity-calibration sequences are completed, and the target charged capacity specified in Section~\ref{sec:fast_charging_generation} is reached during the fast-charging evaluation. Additional screening is applied to \(r_{\mathrm{SOH}}\). In some cases, the simulated 200-cycle SOH loss is very small or slightly negative, and correspondingly, the computed value can become sensitive to solver tolerances. Candidate designs yielding a negative value or numerical zero are excluded because such responses are not informative for learning degradation trends in next section.

Following these checks, \(N=1427\) of the 1501 candidate designs are retained, corresponding to a retention rate of 95.1\%. The resulting quality-controlled simulation dataset is written as
\begin{equation}
\mathcal{D}
=
\left\{
\left(
\mathbf{x}_i, \,
r_{\mathrm{SOH},i}, \, e_{V,i},\,
t_{\mathrm{chg},i}
\right)
\right\}_{i=1}^{N},
\quad
\mathbf{x}_i\in\mathbb{R}^{12}.
\label{eq:quality_controlled_dataset}
\end{equation}

\begin{figure}[htbp]
    \centering
    \includegraphics[width=\linewidth]{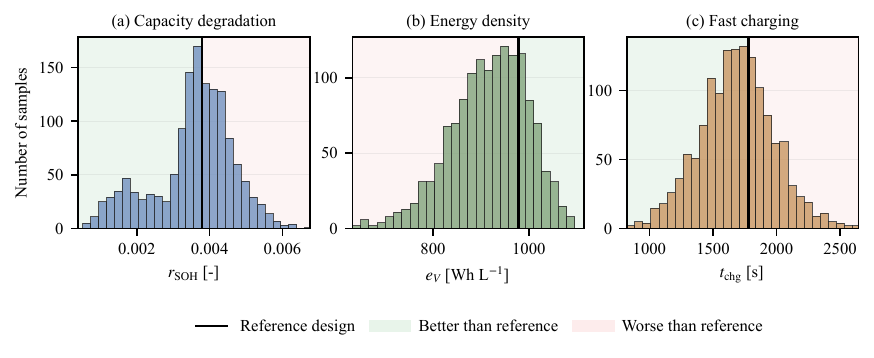}
    \caption{Marginal distributions of the three simulation-derived performance metrics in the quality-controlled dataset (\(N=1427\)). (a) 200-cycle SOH loss \(r_{\mathrm{SOH}}\), (b) stack-level volumetric energy density \(e_V\), and (c) 10--80\% fast-charging time \(t_{\mathrm{chg}}\). The vertical black lines indicate the reference design values.}
    \label{fig:objective_distributions}
\end{figure}

Figure~\ref{fig:objective_distributions} presents the marginal distributions of the three performance metrics across the retained designs. All three distributions extend to both sides of the reference value, indicating that the sampled design space contains designs that outperform the reference in each metric considered separately. The response ranges are substantial: \(r_{\mathrm{SOH}}\) varies from approximately \(3\times10^{-4}\) to \(6.5\times10^{-3}\), \(e_V\) from about 650 to 1100\,Wh\,L\(^{-1}\), and \(t_{\mathrm{chg}}\) from roughly 900 to 2600\,s.

The three metrics exhibit notably different distributional characteristics. The SOH-loss distribution is distinctly non-symmetric, with its dominant concentration close to the reference value and a separate shoulder toward substantially lower degradation. Volumetric energy density is more tightly concentrated, with most designs lying below the reference value but a smaller population extending into the higher-energy-density region. Fast-charging time shows a broad, approximately unimodal distribution around the reference value, with appreciable numbers of both faster- and slower-charging designs. These differences indicate that the 12-dimensional parameter perturbations do not translate uniformly into the three performance responses and provide sufficient output variation for subsequent surrogate-model development. The histograms are marginal, however, and do not show whether lower degradation, higher energy density, and shorter charging time can be achieved simultaneously. This question is addressed later through multi-objective optimization.

\section{Methods}\label{sec:methods}
Joint optimization of battery energy density, fast-charging performance, and lifetime-related performance is challenging for three interconnected reasons. First, the same design variable can affect multiple objectives through competing mechanisms involving active-material loading, transport resistance, reaction heterogeneity, and degradation. Second, although the three objectives are evaluated for the same candidate-design vector, they arise from processes operating over intrinsically different time scales and therefore require objective-specific model fidelities, simulation protocols, and computational horizons. Third, directly embedding the corresponding physics-based models within global sensitivity analysis and evolutionary optimization would require prohibitively expensive computational resources, particularly for the 200-cycle degradation evaluation.

Building on the quality-controlled dataset developed in Section~\ref{sec:data_generation}, the workflow in Fig.~\ref{fig:workflow} contains three computational components followed by physics-based confirmation. Objective-specific surrogate models first replace repeated calls to the corresponding electrochemical simulations. Total-order Sobol indices and the rank-union strategy then identify reduced variable sets that retain parameters important to any objective. NSGA-II subsequently searches a sequence of nested, sensitivity-informed design spaces. The resulting Pareto candidates are finally re-evaluated using the original multiphysics battery model to confirm numerical feasibility and determine their actual improvements in the three performance metrics.

\subsection{Surrogate model construction}
\label{subsec:surrogate_model_construction}
Surrogate-assisted optimization has proved effective for battery-manufacturing descriptors, energy--power design, fast-charging co-design, and degradation-aware optimization~\cite{Duquesnoy2023multiobjective,duquesnoy2024highperformance,cui2022codesign,ma2024multiobjective,ju2026multiobjective}. Its role in the present problem is particularly important because the cost of generating the three targets is highly uneven. A BoL energy-density or charging simulation is inexpensive relative to repeated aging cycles, yet all three models must be called many times during the subsequent sensitivity analysis and Pareto search. The surrogate construction therefore retains alignment where it is physically justified---in the common design space---while allowing the response model to differ among objectives.

As described in Section~\ref{sec:data_generation}, \(e_V\) and \(t_{\mathrm{chg}}\) are obtained from BoL DFN simulations, whereas \(r_{\mathrm{SOH}}\) is calculated from the 200-cycle SPMe degradation simulations. The quality-controlled dataset in Eq.~\eqref{eq:quality_controlled_dataset} is accordingly decomposed into three aligned, objective-specific regression datasets:
\begin{equation}
\mathcal{D}_{q} =
\left\{
\left(\mathbf{x}_i,\, y^{(q)}_{i}\right)
\right\}^{N}_{i=1},
\quad
q\in\left\{\mathrm{SOH},E,t\right\},
\label{eq:objective_datasets}
\end{equation}
where
\begin{equation}
y_{i}^{(\mathrm{SOH})}=r_{\mathrm{SOH},\,i},
\quad
y_{i}^{(E)}=e_{V,\,i},
\quad
y_{i}^{(t)}=t_{\mathrm{chg},\,i}.
\label{eq:objective_targets}
\end{equation}
The datasets contain the same candidate-design vectors but different scalar targets. They are therefore aligned in design space, not in physical time scale or simulation protocol. Separate training and model-selection procedures are used because the three targets exhibit different distributions, response-surface geometries, sensitivity structures, and levels of predictability.

For candidate design \(\mathbf{x}_{i}\), the surrogate predictions are expressed as
\begin{equation}
\hat{r}_{\mathrm{SOH},\,i}^{(M)} =
\hat{g}_{\mathrm{SOH}}^{(M)}(\mathbf{x}_{i}),
\qquad
\hat{e}_{V,\,i}^{(M)} =
\hat{g}_{E}^{(M)}(\mathbf{x}_{i}),
\qquad
\hat{t}_{\mathrm{chg},\,i}^{\,(M)} =
\hat{g}_{t}^{(M)}(\mathbf{x}_{i}),
\label{eq:surrogate_models}
\end{equation}
where \(M\) denotes a candidate machine-learning method.

Three machine-learning methods are employed and represent complementary regression structures. Namely, in Eq.~\eqref{eq:surrogate_models}, \(M\in\{\mathrm{RFR, HGBR, SVR}\}\).
Random forest regression (RFR) averages an ensemble of decorrelated decision trees generated through bootstrap resampling and random feature selection~\cite{breiman2001random}. It can represent nonlinear thresholds and interactions among geometric, material, and transport parameters without prescribing a functional form. Histogram-based gradient boosting regression (HGBR) instead constructs an additive tree ensemble sequentially, with each tree reducing the residual error of the preceding ensemble~\cite{friedman2001greedy}. Histogram binning reduces the computational cost of repeated split evaluation. Support vector regression (SVR) with a radial basis function (RBF)-based kernel provides a regularized, non-tree-based alternative in which a smooth nonlinear mapping is learned through kernel similarity~\cite{smola2004tutorial}. Comparing bagged trees, boosted trees, and kernel regression avoids imposing the same approximation structure on three physically different response surfaces.

Each row of \(\mathcal{D}_{q}\) represents one candidate cell design rather than one point in a time sequence. The 200-cycle degradation trajectory has already been condensed into the scalar target \(r_{\mathrm{SOH}}\), and no temporal ordering is imposed during regression. The LHS realizations are finite, but the underlying design variables remain continuous within their prescribed bounds, and nonlinear interactions among them are retained by all three candidate model families.
For every combination of objective \(q\) and regression method \(M\), the hyperparameters are tuned independently using five-fold cross-validation. Predictions for each design are obtained from a model that excludes that design from its training fold. Predictive accuracy is evaluated using the coefficient of determination,
\begin{equation}
R^{2} =
1-
\frac{
\sum_{i=1}^{N}
\left(y_i-\hat{y}_i\right)^2
}{
\sum_{i=1}^{N}
\left(y_i-\bar{y}\right)^2
},
\label{eq:surrogate_r2}
\end{equation}
and the relative mean absolute error,
\begin{equation}
\mathrm{RMAE} =
\frac{
\sum_{i=1}^{N}
\left|y_i-\hat{y}_i\right|
}{
N \,\bar{y}
}
\times 100\%,
\label{eq:surrogate_rmae}
\end{equation}
where \(y_i\), \(\hat{y}_i\), and \(\bar{y}\) denote the simulation-derived target, its cross-validated prediction, and the mean target value, respectively. Higher \(R^{2}\) and lower \(\mathrm{RMAE}\) indicate stronger predictive performance. The selected model family and hyperparameters can therefore differ among the three objectives. After model selection, the chosen surrogate is refitted using the complete objective-specific dataset. 

\subsection{Sobol-guided optimization-variable selection}
\label{subsec:sobol_rank_union}

Although the surrogate models substantially reduce the cost of evaluating the three objectives, simultaneously optimizing all 12 design parameters still creates a high-dimensional continuous search problem. For a fixed computational budget, the effective coverage of the design space decreases rapidly as the number of adjustable parameters increases. The NSGA-II population then samples the design space more sparsely, slowing convergence and increasing the risk that relevant regions of a nonlinear and nonconvex Pareto set remain insufficiently explored. Parameters with negligible influence on all three objectives also consume search degrees of freedom without providing meaningful optimization leverage. Sensitivity analysis is therefore used to construct reduced optimization problems.

The screening criterion must capture nonlinear responses and interactions across the prescribed design space. Pearson and Spearman correlation coefficients could be calculated directly from the simulation dataset in Section~\ref{sec:data_generation}, but they quantify marginal linear and monotonic associations, respectively~\cite{zhu2024predicting}. A parameter whose influence is non-monotonic or arises primarily through interactions may consequently exhibit weak correlation despite being important to the objective response. We instead apply variance-based Sobol analysis to the objective-specific surrogate models selected in Section~\ref{subsec:surrogate_model_construction}. Sobol analysis attributes the predicted output variance to individual inputs and their interactions over the prescribed input domain~\cite{sobol2001global,saltelli2010variance}.

For the sensitivity analysis, the 12 design parameters are represented by the random vector \(\mathbf{X}=[X_1,\ldots,X_{12}]^{\mathsf T}\), where
\begin{equation}
X_j
\sim
\mathcal{U}
\left(
0.8x_{j,0},
1.2x_{j,0}
\right),
\quad
j=1,\ldots,12,
\label{eq:sobol_input_distribution}
\end{equation}
and the inputs are treated as mutually independent. The uniform distributions define the probability measure used for sensitivity analysis over the computational design space introduced in Section~\ref{sec:design_space_sampling}. Here, ``global'' sensitivity refers to integration over the complete prescribed \(\pm20\%\) domain around the reference design, rather than over an unrestricted range of physically possible cell designs.

Let \(g^{(q)}\) denote the selected surrogate model for objective
\(q\), refitted using the complete objective-specific dataset. The first-order Sobol index measures the output variance attributable to the main effect of one parameter. The total-order index additionally includes all interaction effects involving that parameter. Because electrode geometry, active-material capacity, solid-state diffusion, and electrolyte transport are coupled in the underlying cell models, parameter ranking is based on the total-order index. For parameter \(X_j\) and objective \(q\), it is defined as
\begin{equation}
S_{T,j}^{(q)}
=
\frac{
\mathbb{E}_{\mathbf{X}_{\sim j}}
\left[
\operatorname{Var}_{X_j}
\left(
g^{(q)}(\mathbf{X})
\mid
\mathbf{X}_{\sim j}
\right)
\right]
}{
\operatorname{Var}_{\mathbf{X}}
\left(
g^{(q)}(\mathbf{X})
\right)
},
\label{eq:total_order_sobol}
\end{equation}
where \(\mathbf{X}_{\sim j}\) contains all input parameters except \(X_j\). For a fixed realization of \(\mathbf{X}_{\sim j}\), the conditional variance in the numerator measures how much \(g^{(q)}(\mathbf{X})\) varies when \(X_j\) alone is varied over its prescribed range. The outer expectation averages this conditional variance over the possible configurations of the other 11 parameters. The denominator is the total variance of \(g^{(q)}(\mathbf{X})\) when all 12 parameters vary. The resulting ratio therefore captures both the main effect of \(X_j\) and every interaction involving \(X_j\).

Equation~\eqref{eq:total_order_sobol} defines a population quantity over the input distributions in Eq.~\eqref{eq:sobol_input_distribution}. Numerically, its expectations and variances are estimated from a structured Saltelli sampling ensemble, denoted by
\begin{equation}
\mathcal{X}_{\mathrm{SA}}
=
\left\{
\mathbf{X}^{(s)}
\right\}_{s=1}^{N_{\mathrm{eval}}}
=
\left\{
\mathbf{X}^{(1)},\,
\mathbf{X}^{(2)},\,
\ldots,\,
\mathbf{X}^{(N_{\mathrm{eval}})}
\right\},
\label{eq:saltelli_sampling_ensemble}
\end{equation}
where \(s\) denotes the sensitivity-analysis evaluation index. For the adopted Saltelli sampling configuration, a base sample size of \(N_{\mathrm{base}}=4096\) produces
\begin{equation}
N_{\mathrm{eval}}
=
(J+2)\,N_{\mathrm{base}}
=
57\,344,
\quad
J=12.
\label{eq:saltelli_evaluation_number}
\end{equation}
The surrogate responses \(g^{(q)}(\mathbf{X}^{(s)})\), \(s=1,\ldots,N_{\mathrm{eval}}\), are then used to estimate the expectations and variances in Eq.~\eqref{eq:total_order_sobol}. The index \(s\) is distinct from the candidate-design index \(i\) used to construct the regression datasets, and the Saltelli ensemble is generated separately from the candidate designs used to train the surrogate models.

The theoretical value of \(S_{T,j}^{(q)}\) lies between 0 and 1. Total-order indices need not sum to one because an interaction contribution is included in the index of every parameter participating in that interaction. A value close to zero indicates that varying \(X_j\) contributes little to the predicted objective variance within the prescribed design space. A larger value indicates stronger influence, but not the direction of that influence; it does not imply that increasing or decreasing \(X_j\) improves the corresponding objective. Moreover, because the indices are calculated from surrogate predictions, they characterize the fitted response surfaces over the stated \(\pm20\%\) domain rather than constituting domain-independent sensitivity properties of the original physics-based models.

For each objective \(q\), let
\(\boldsymbol{\pi}_q=
[\pi_{q,1},\ldots,\pi_{q,12}]\)
denote a permutation of the parameter indices satisfying
\begin{equation}
S_{T,\,\pi_{q,1}}^{(q)}
\geq
S_{T,\,\pi_{q,2}}^{(q)}
\geq
\cdots
\geq
S_{T,\,\pi_{q,12}}^{(q)}.
\label{eq:objective_specific_ranking}
\end{equation}
The corresponding objective-specific ranked parameter list is
\begin{equation}
\mathcal{R}_{q}
=
\left[
x_{\pi_{q,1}},
x_{\pi_{q,2}},
\ldots,
x_{\pi_{q,12}}
\right].
\label{eq:objective_specific_ranked_list}
\end{equation}
The rankings are constructed separately because a parameter with little influence on one objective may remain important for another.

For rank level \(k\), the top-\(k\) parameter set for objective \(q\) is defined as
\begin{equation}
\mathcal{S}_{q,k}
=
\left\{
x_{\pi_{q,1}},
x_{\pi_{q,2}},
\ldots,
x_{\pi_{q,k}}
\right\}.
\label{eq:top_k_parameter_set}
\end{equation}
The corresponding rank-union optimization-variable set is
\begin{equation}
\mathcal{P}_{k}
=
\mathcal{S}_{\mathrm{SOH},k}
\cup
\mathcal{S}_{E,k}
\cup
\mathcal{S}_{t,k}.
\label{eq:rank_union_parameter_set}
\end{equation}
Because the set union removes duplicate parameters, \(\mathcal{P}_{k}\) contains every parameter appearing within the top \(k\) positions for at least one objective. A parameter is excluded only when it lies outside the top \(k\) positions for all three objectives. This conservative cross-objective construction prevents the dimension-reduction procedure from being governed by a single performance metric.

Since
\(\mathcal{S}_{q,k}\subseteq\mathcal{S}_{q,k+1}\),
the rank-union sets are nested:
\begin{equation}
\mathcal{P}_{k}
\subseteq
\mathcal{P}_{k+1}.
\label{eq:nested_rank_union_sets}
\end{equation}
Their cardinality need not increase at every rank level because the objective-specific rankings can overlap. In the optimization described in Section~\ref{subsec:nsga2_optimization}, only the parameters contained in a given \(\mathcal{P}_{k}\) are allowed to vary; all parameters outside that set are assigned their reference values \(x_{j,0}\).

\subsection{Surrogate-based Pareto optimization}
\label{subsec:nsga2_optimization}

For each investigated rank level \(k\), the three objective-specific surrogate models selected in Section~\ref{subsec:surrogate_model_construction} are optimized over the corresponding rank-union variable set \(\mathcal{P}_{k}\) constructed in Section~\ref{subsec:sobol_rank_union}. The rank level is retained as a tuning parameter: changing \(k\) alters the number of adjustable design parameters without changing the objective definitions. Let
\[
\mathcal{J}_{k}
=
\left\{
j\in\{1,\ldots,12\}:x_j\in\mathcal{P}_{k}
\right\}
\]
denote the indices of the parameters included in \(\mathcal{P}_{k}\). The reduced multi-objective problem at rank level \(k\) is formulated as
\begin{equation}
\begin{aligned}
\underset{\mathbf{x}\in\mathbb{R}^{12}}{\operatorname{minimize}}
\quad
&\mathbf{g}(\mathbf{x})
=
\begin{bmatrix}
g^{(\mathrm{SOH})}(\mathbf{x})\\[2pt]
-g^{(E)}(\mathbf{x})\\[2pt]
g^{(t)}(\mathbf{x})
\end{bmatrix},
\\
\text{subject to}
\quad
&0.8x_{j,0}
\leq
x_j
\leq
1.2x_{j,0},
\quad
j\in\mathcal{J}_{k},
\\
&x_j=x_{j,0},
\quad
j\in\{1,\ldots,12\}\setminus\mathcal{J}_{k}.
\end{aligned}
\label{eq:reduced_multiobjective_problem}
\end{equation}
Here, \(\mathbf{x}=[x_1,\ldots,x_{12}]^{\mathsf T}\) denotes a candidate cell design. 
The three components of \(\mathbf{g}\) are supplied directly by the selected and refitted surrogate models \(g^{(\mathrm{SOH})}\), \(g^{(E)}\), and \(g^{(t)}\). The negative sign converts volumetric energy-density maximization into minimization, allowing all three objectives to be treated in a common direction. Eq.~\eqref{eq:reduced_multiobjective_problem} is interpreted in the Pareto sense; its solution is generally a set of mutually non-dominated designs rather than a single optimal parameter vector.

The reduced problems are solved using NSGA-II~\cite{deb2002fast}. This choice is well matched to the present formulation. The independently selected surrogate models can produce nonlinear, nonconvex, and potentially non-smooth response surfaces, for which reliable analytical gradients are not generally available. NSGA-II does not require such gradients or an a priori scalarization of the three objectives. Its elitist non-dominated sorting promotes convergence toward the Pareto front, whereas crowding-distance selection maintains coverage across different regions of the trade-off surface.

Although variation is restricted to the coordinates indexed by \(\mathcal{J}_{k}\), every candidate is represented as a complete 12-dimensional design vector when evaluated by the surrogate models. The NSGA-II procedure used for each rank level  \(k\) is summarized in Algorithm~\ref{alg:nsga2_reduced}. The population size and number of generations are denoted by \(N_{\mathrm{pop}}\) and \(N_{\mathrm{gen}}\), respectively. The notation \(\mathcal{F}_1(\mathcal{G}_{N_{\mathrm{gen}}})\) denotes the first non-dominated front of the final population \(\mathcal{G}_{N_{\mathrm{gen}}}\). 

\begin{algorithm}[ht]
\caption{NSGA-II search for the rank-\(k\) reduced multi-objective problem}
\label{alg:nsga2_reduced}
\begin{algorithmic}[1]
\Require Rank level \(k\), index set \(\mathcal{J}_{k}\), reference design \(\mathbf{x}_0\), and surrogate models \(g^{(\mathrm{SOH})}\), \(g^{(E)}\), and \(g^{(t)}\)
\Require Population size \(N_{\mathrm{pop}}\), number of generations \(N_{\mathrm{gen}}\), and crossover and mutation settings
\Ensure First non-dominated front of the final population

\State Randomly generate an initial population \(\mathcal{G}_0\) of \(N_{\mathrm{pop}}\) designs satisfying Eq.~\eqref{eq:reduced_multiobjective_problem}
\State Evaluate \(\mathbf{g}(\mathbf{x})\) for all \(\mathbf{x}\in\mathcal{G}_0\)
\State Assign non-domination ranks and crowding distances to the members of \(\mathcal{G}_0\)

\ForLoop{\(\ell=0,\ldots,N_{\mathrm{gen}}-1\)}
    \State Select parent designs from \(\mathcal{G}_{\ell}\) using crowded-comparison tournament selection
    \State Apply crossover and mutation only to the coordinates indexed by \(\mathcal{J}_{k}\)
    \State Enforce the prescribed bounds and set \(x_j=x_{j,0}\) for every \(j\notin\mathcal{J}_{k}\)
    \State Form an offspring population \(\mathcal{Q}_{\ell}\) of size \(N_{\mathrm{pop}}\)
    \State Evaluate \(\mathbf{g}(\mathbf{x})\) for all \(\mathbf{x}\in\mathcal{Q}_{\ell}\)
    \State Combine the populations:
    \[
    \mathcal{R}_{\ell}
    =
    \mathcal{G}_{\ell}
    \cup
    \mathcal{Q}_{\ell}
    \]
    \State Partition \(\mathcal{R}_{\ell}\) into non-dominated fronts and compute the crowding distance within each front
    \State Add complete fronts to \(\mathcal{G}_{\ell+1}\) in ascending rank order
    \State If the next front cannot be retained in full, select its members in descending order of crowding distance until \(|\mathcal{G}_{\ell+1}|=N_{\mathrm{pop}}\)
\EndForLoop

\State \Return \(
\mathcal{F}_1\!\left(\mathcal{G}_{N_{\mathrm{gen}}}\right)
\)

\end{algorithmic}
\end{algorithm}

Non-dominated sorting partitions the combined parent--offspring population into successive Pareto fronts. Candidates in the first front are not dominated by any other member of the combined population; subsequent fronts are identified after the preceding fronts have been removed. Crowding distance is used only to distinguish candidates within the same front. A larger distance indicates a less densely represented region of the objective space and therefore receives priority when a front cannot be retained in full. Combining parent and offspring populations before selection gives the procedure its elitist character: previously identified non-dominated candidates are not discarded merely because they were generated in an earlier generation.

After \(N_{\mathrm{gen}}\) generations, the first non-dominated front of the final population is retained as the surrogate-derived Pareto candidate set for the corresponding \(\mathcal{P}_{k}\). These candidates remain optimal only with respect to the fitted surrogate responses and the rank-\(k\) reduced design space. They are therefore re-evaluated using the original physics-based models before the final performance gains are assessed.

\section{Results and discussion}
\subsection{Evaluation of surrogate model accuracy}
\label{subsec:surrogate_accuracy}

The Sobol-based parameter ranking and Pareto optimization rely on the
accuracy of the surrogate response surfaces. The candidate predictors
\(\hat{g}_{q}^{(M)}\) introduced in Eq.~\eqref{eq:surrogate_models}
are therefore evaluated separately for
\(q\in\{\mathrm{SOH},E,t\}\) and
\(M\in\{\mathrm{RFR},\mathrm{HGBR},\mathrm{SVR}\}\).
For each objective--model pair, the hyperparameters are tuned independently using the five-fold cross-validation procedure described in
Section~\ref{subsec:surrogate_model_construction}. The resulting settings are reported
in Table~\ref{tab:tuned_model_hyperparameters}, and the corresponding
cross-validated \(R^2\) and RMAE values are summarized in
Table~\ref{tab:surrogate_performance}.

\begin{table*}[ht]
\centering
\caption{Hyperparameter settings selected by five-fold cross-validation
for the RFR, HGBR, and SVR candidate surrogate models.}
\label{tab:tuned_model_hyperparameters}
\small
\setlength{\tabcolsep}{4pt}
\renewcommand{\arraystretch}{1.05}

\begin{tabular*}{\textwidth}{@{\extracolsep{\fill}}lccccc@{}}
\hline
\multicolumn{6}{c}{\textbf{RFR}} \\
\hline
\rule[-0.8ex]{0pt}{5.4ex}%
\shortstack[l]{Performance\\metric}
& \shortstack{Number of\\trees}
& \multicolumn{2}{c}{\shortstack{Features considered\\at each split}}
& \multicolumn{2}{c}{\shortstack{Minimum samples\\per leaf}} \\
\hline
\(r_{\mathrm{SOH}}\)
& 600
& \multicolumn{2}{c}{80\% of the input features}
& \multicolumn{2}{c}{2} \\

\(e_V\)
& 600
& \multicolumn{2}{c}{80\% of the input features}
& \multicolumn{2}{c}{1} \\

\(t_{\mathrm{chg}}\)
& 600
& \multicolumn{2}{c}{Square root of the input-feature count}
& \multicolumn{2}{c}{1} \\
\hline

\multicolumn{6}{c}{\textbf{HGBR}} \\
\hline
\rule[-0.8ex]{0pt}{5.4ex}%
\shortstack[l]{Performance\\metric}
& \shortstack{Maximum boosting\\iterations}
& \shortstack{Learning\\rate} 
& \shortstack{Maximum leaf\\nodes}
& \shortstack{Minimum samples\\per leaf}
& \shortstack{\(L_2\)-regularization\\strength} \\
\hline
\(r_{\mathrm{SOH}}\)
& 1000 & 0.04 & 31 & 10 & \(1\times10^{-4}\) \\
\(e_V\)
& 1000 & 0.06 & 15 & 40 & \(1\times10^{-5}\) \\

\(t_{\mathrm{chg}}\)
& 1000 & 0.04 & 15 & 40 & \(1\times10^{-4}\) \\
\hline

\multicolumn{6}{c}{\textbf{SVR}} \\
\hline
\rule[-0.8ex]{0pt}{5.4ex}%
\shortstack[l]{Performance\\metric}
& \shortstack{Regularization\\penalty}
& \multicolumn{2}{c}{\shortstack{Insensitive-region\\width}}
& \multicolumn{2}{c}{\shortstack{RBF-kernel\\coefficient}} \\
\hline
\(r_{\mathrm{SOH}}\)
& 10
& \multicolumn{2}{c}{0.10}
& \multicolumn{2}{c}{0.03} \\

\(e_V\)
& 30
& \multicolumn{2}{c}{0.03}
& \multicolumn{2}{c}{0.03} \\

\(t_{\mathrm{chg}}\)
& 30
& \multicolumn{2}{c}{0.03}
& \multicolumn{2}{c}{0.03} \\
\hline
\end{tabular*}
\end{table*}

\begin{table}[ht]
\centering
\caption{Five-fold cross-validation performance of the tuned candidate surrogate models on the quality-controlled dataset.}
\label{tab:surrogate_performance}
\begin{tabular}{llcc}
\hline
Performance metric & Model & \(R^2\) & RMAE [\%] \\
\hline
\(r_{\mathrm{SOH}}\)
& RFR  & 0.714 & 10.54 \\
& HGBR & \textbf{0.770} & \textbf{9.81} \\
& SVR  & 0.735 & 11.15 \\
\hline
\(e_V\)
& RFR  & 0.769 & 3.20 \\
& HGBR & 0.889 & 2.22 \\
& SVR  & \textbf{0.927} & \textbf{1.77} \\
\hline
\(t_{\mathrm{chg}}\)
& RFR  & 0.762 & 6.51 \\
& HGBR & 0.907 & 3.99 \\
& SVR  & \textbf{0.920} & \textbf{3.68} \\
\hline
\end{tabular}
\end{table}

The preferred regression structure is objective dependent. HGBR provides the most accurate prediction of \(r_{\mathrm{SOH}}\), with \(R^2=0.770\) and an RMAE of \(9.81\%\), whereas SVR performs best for both \(e_V\) and \(t_{\mathrm{chg}}\). For every performance metric, the model with the lowest RMAE also attains the highest \(R^2\), making the objective-specific selection unambiguous. Accordingly, in Eqs.~\eqref{eq:total_order_sobol} and~\eqref{eq:reduced_multiobjective_problem}, \(g^{(\mathrm{SOH})}\) is obtained from the tuned HGBR predictor, while \(g^{(E)}\) and \(g^{(t)}\) are obtained from their respective tuned SVR predictors. Each selected model is subsequently refitted to the complete objective-specific dataset with its hyperparameters held fixed.

Among the three targets, \(e_V\) presents the most readily approximated response surface: the selected SVR explains \(92.7\%\) of its cross-validated variance with an RMAE of only \(1.77\%\). Within the prescribed design space, stack-level energy density depends relatively directly on electrode loading, active-material storage capacity, and stack thickness, producing a smooth mapping that is well represented by the RBF kernel. The \(t_{\mathrm{chg}}\) surrogate also achieves high predictive accuracy, with \(R^2=0.920\) and an RMAE of \(3.68\%\). Under the fixed-current protocol adopted here, \(t_{\mathrm{chg}}\) is closely linked to \(Q_{\mathrm{init}}\) for designs that reach the prescribed charge throughput. The spatial electrochemical dynamics resolved by the DFN model primarily determine whether the upper-voltage cut-off is reached before that throughput is delivered.

The degradation response is appreciably more difficult to approximate. Unlike the two BoL metrics, \(r_{\mathrm{SOH}}\) accumulates the coupled effects of three degradation mechanisms over 200 cycles. It is also evaluated from the relatively small difference between two calibrated capacities, \(Q_{\mathrm{init}}\) and \(Q_{\mathrm{final}}\), and exhibits the strongly skewed distribution shown in Fig.~\ref{fig:objective_distributions}. These features produce a more locally varying response surface and make small degradation values particularly demanding to learn. Consistent with this behavior, the tuned HGBR model uses finer leaf partitions than those selected for \(e_V\) and \(t_{\mathrm{chg}}\). Despite the more complex degradation response, the selected model maintains a cross-validated RMAE below \(10\%\) for \(r_{\mathrm{SOH}}\) and is retained as the degradation surrogate for the subsequent sensitivity analysis and optimization within the prescribed design domain.


\subsection{Sobol-guided optimization-variable sets}
\label{subsec:selected_optimization_variables}

Using the objective-specific surrogate models selected in Section~\ref{subsec:surrogate_accuracy}, we evaluate the total-order Sobol indices of the 12 design parameters following the procedure in Section~\ref{subsec:sobol_rank_union}. Table~\ref{tab:sobol_total_indices} reports the resulting indices, and Table~\ref{tab:rank_union_variable_sets} gives the reduced parameter sets carried forward to multi-objective optimization.

\begin{table}[!htbp]
\centering
\caption{Sobol total-order sensitivity indices of the 12 design parameters for the selected surrogate models of the three performance metrics. The parameters are presented in descending order of \(S_{T,j}^{(\mathrm{SOH})}\). Each \(S_{T,j}^{(q)}\) measures the contribution of parameter \(j\) to the surrogate-predicted variance of metric \(q\), including all interaction effects involving that parameter. The three largest indices for each performance metric are highlighted in bold.}
\label{tab:sobol_total_indices}
\begin{tabular}{lccc}
\hline
Parameter & \(r_{\mathrm{SOH}}\) & \(e_V\) & \(t_{\mathrm{chg}}\) \\
\hline
\(R_{\mathrm{n}}\)             & \(\mathbf{0.459}\) & \(0.010\)          & \(0.008\) \\
\(c_{\mathrm{p,max}}\)         & \(\mathbf{0.230}\) & \(\mathbf{0.486}\) & \(\mathbf{0.382}\) \\
\(L_{\mathrm{n}}\)             & \(\mathbf{0.216}\) & \(\mathbf{0.256}\) & \(0.064\) \\
\(c_{\mathrm{n,max}}\)         & \(0.209\)          & \(0.194\)          & \(\mathbf{0.120}\) \\
\(R_{\mathrm{p}}\)             & \(0.096\)          & \(0.057\)          & \(0.079\) \\
\(L_{\mathrm{p}}\)             & \(0.086\)          & \(\mathbf{0.249}\) & \(\mathbf{0.302}\) \\
\(\varepsilon_{\mathrm{n}}\)   & \(0.038\)          & \(0.113\)          & \(0.063\) \\
\(\varepsilon_{\mathrm{p}}\)   & \(0.003\)          & \(0.118\)          & \(0.111\) \\
\(b_{\mathrm{s}}\)             & \(0.002\)          & \(0.004\)          & \(0.006\) \\
\(c_{\mathrm{e},0}\)           & \(0.002\)          & \(0.005\)          & \(0.009\) \\
\(b_{\mathrm{n}}\)             & \(0.002\)          & \(0.005\)          & \(0.073\) \\
\(b_{\mathrm{p}}\)             & \(0.002\)          & \(0.005\)          & \(0.023\) \\
\hline
\end{tabular}
\end{table}

The sensitivity is concentrated in a small group of electrode-level parameters, but the ranking remains specific to the performance metric. The only parameter ranked among the three most influential for every metric is \(c_{\mathrm{p,max}}\), with total-order indices of \(0.230\), \(0.486\), and \(0.382\) for \(r_{\mathrm{SOH}}\), \(e_V\), and \(t_{\mathrm{chg}}\), respectively. By setting the positive-electrode lithium-storage capacity, \(c_{\mathrm{p,max}}\) affects usable cell capacity and the capacity balance between the two electrodes. Shared sensitivity is also evident in the electrode thicknesses: \(L_{\mathrm{n}}\) ranks among the top three for \(r_{\mathrm{SOH}}\) and \(e_V\), whereas \(L_{\mathrm{p}}\) does so for \(e_V\) and \(t_{\mathrm{chg}}\). These results identify where variations in loading, stack geometry, and transport length enter more than one performance response. They do not, however, imply that changing a shared parameter will improve all three metrics in the same direction, because total-order indices contain no directional information.

The differences among the rankings are equally consequential. \(R_{\mathrm{n}}\) dominates the variance of the predicted SOH loss, with \(S_{T,j}^{(\mathrm{SOH})}=0.459\), but contributes little to the predicted variances of \(e_V\) and \(t_{\mathrm{chg}}\), for which its indices are \(0.010\) and \(0.008\). Within the adopted SEI-growth and stress-driven loss-of-active-material model, this degradation sensitivity is consistent with the effects of \(R_{\mathrm{n}}\) on specific interfacial area, solid-state diffusion length, and particle stress. In contrast, \(L_{\mathrm{p}}\) strongly influences \(e_V\) and \(t_{\mathrm{chg}}\) but has a more moderate index for \(r_{\mathrm{SOH}}\). Screening against degradation alone would therefore under-prioritize \(L_{\mathrm{p}}\), whereas screening against energy density or charging time would largely overlook \(R_{\mathrm{n}}\). Our rank-union construction in Eq.~\eqref{eq:rank_union_parameter_set} prevents either outcome.

Parameter interactions are evident in all three fitted response surfaces. The sums \(\sum_{j=1}^{12} S_{T,j}^{(q)}\) are approximately \(1.35\), \(1.50\), and \(1.24\) for \(r_{\mathrm{SOH}}\), \(e_V\), and \(t_{\mathrm{chg}}\), respectively. The excess above unity reflects the multiple counting of interaction contributions across the total-order indices. At the low-sensitivity end, \(b_{\mathrm{s}}\) and \(c_{\mathrm{e},0}\) have total-order indices below \(0.01\) for every metric. Within the prescribed parameter ranges and operating protocols, these two parameters account for little of the surrogate-predicted variance. This screening result is specific to the investigated design domain and does not imply that separator transport or electrolyte concentration is generally unimportant.

\begin{table}[!htbp]
\centering
\caption{Nested Sobol rank-union sets used for multi-objective optimization. At rank level \(k\), the top-\(k\) parameters from the three metric-specific Sobol rankings are combined and duplicate parameters are removed. The unions at rank levels 6, 7, and 8 are identical; their common nine-parameter set is denoted by \(\mathcal{P}_6\).}
\label{tab:rank_union_variable_sets}
\begin{tabular}{llcl}
\hline
Set & Rank-union rule & No. of parameters & Adjustable design parameters \\
\hline
\(\mathcal{P}_1\) & Rank-1 union & 2
& \(R_{\mathrm{n}}, c_{\mathrm{p,max}}\) \\
\(\mathcal{P}_2\) & Rank-2 union & 4
& \(R_{\mathrm{n}}, c_{\mathrm{p,max}}, L_{\mathrm{n}}, L_{\mathrm{p}}\) \\
\(\mathcal{P}_3\) & Rank-3 union & 5
& \(R_{\mathrm{n}}, c_{\mathrm{p,max}}, L_{\mathrm{n}}, L_{\mathrm{p}}, c_{\mathrm{n,max}}\) \\
\(\mathcal{P}_4\) & Rank-4 union & 6
& \(R_{\mathrm{n}}, c_{\mathrm{p,max}}, L_{\mathrm{n}}, L_{\mathrm{p}}, c_{\mathrm{n,max}}, \varepsilon_{\mathrm{p}}\) \\
\(\mathcal{P}_5\) & Rank-5 union & 7
& \(R_{\mathrm{n}}, c_{\mathrm{p,max}}, L_{\mathrm{n}}, L_{\mathrm{p}}, c_{\mathrm{n,max}}, \varepsilon_{\mathrm{p}}, R_{\mathrm{p}}\) \\
\(\mathcal{P}_6\) & Rank-6--8 union & 9
& \(R_{\mathrm{n}}, c_{\mathrm{p,max}}, L_{\mathrm{n}}, L_{\mathrm{p}}, c_{\mathrm{n,max}}, \varepsilon_{\mathrm{p}}, R_{\mathrm{p}}, \varepsilon_{\mathrm{n}}, b_{\mathrm{n}}\) \\
\hline
\end{tabular}
\end{table}

The rank-union procedure converts these metric-specific rankings into six nested optimization-variable sets, \(\mathcal{P}_1 \subseteq \mathcal{P}_2 \subseteq \cdots \subseteq \mathcal{P}_6\). Their sizes do not increase uniformly with rank level because a parameter selected for one metric may already have been selected for another. At Rank~1, \(c_{\mathrm{p,max}}\) leads the rankings for both \(e_V\) and \(t_{\mathrm{chg}}\), while \(R_{\mathrm{n}}\) leads the ranking for \(r_{\mathrm{SOH}}\); \(\mathcal{P}_1\) therefore contains two parameters rather than three. By Rank~3, the union contains five parameters. Increasing the rank level to 6 adds \(\varepsilon_{\mathrm{p}}\), \(R_{\mathrm{p}}\), \(\varepsilon_{\mathrm{n}}\), and \(b_{\mathrm{n}}\), each of which appears within the top six for at least one metric.

The Rank-6, Rank-7, and Rank-8 unions yield the same nine-parameter set, represented by \(\mathcal{P}_6\). The remaining parameters, \(b_{\mathrm{s}}\), \(c_{\mathrm{e},0}\), and \(b_{\mathrm{p}}\), stay fixed at their reference values; their largest total-order indices across the three metrics are \(0.006\), \(0.009\), and \(0.023\), respectively. We therefore stop the sequence before Rank~9. The resulting sets span the practical trade-off between optimization dimensionality and cell performance: a smaller adjustable set permits denser exploration under a fixed NSGA-II evaluation budget, whereas releasing additional parameters enlarges the search space but may improve the attainable Pareto solutions. Evaluating \(\mathcal{P}_1\)--\(\mathcal{P}_6\) allows this trade-off to be quantified through subsequent physics-based validation.

\subsection{Results of surrogate-assisted Pareto optimization}
\label{subsec:optimization_validation_results}

\begin{figure}[ht]
    \centering
    \includegraphics[width=\linewidth]{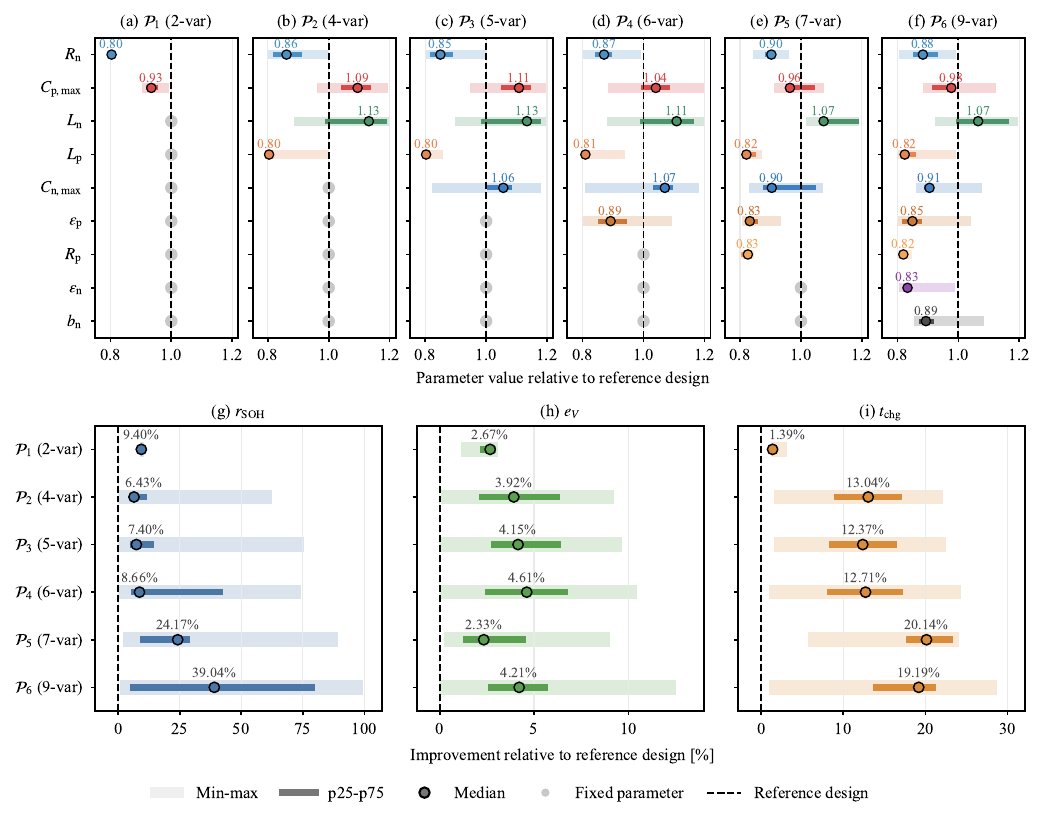}
    \caption{Validated jointly improving designs across the six rank-union parameter sets. Panels (a)--(f) show reference-normalized parameter distributions of candidates that outperform the reference design in all three objectives for \(\mathcal{P}_1\)--\(\mathcal{P}_6\). Light and dark bars denote the min--max and 25--75th percentile ranges, respectively, circles denote medians, and gray points denote fixed parameters. Panels (g)--(i) show the corresponding improvement distributions of SOH loss (\(\Delta_{\mathrm{SOH}}\)), volumetric energy density (\(\Delta_E\)), and fast-charging time (\(\Delta_t\)).}
    \label{fig:strict_pass_improvement_distributions}
\end{figure}

\begin{figure}[ht]
    \centering
    \includegraphics[width=0.98\linewidth]{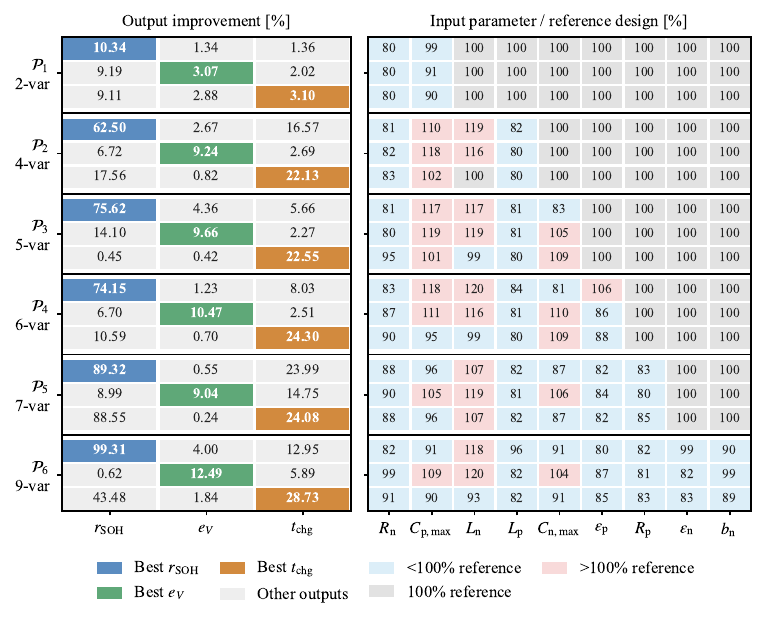}
    \caption{Objective-specific best designs among the validated jointly improving candidates. For each reduced variable set \(\mathcal{P}_k\), the three rows select the largest \(\Delta_{\mathrm{SOH}}\), \(\Delta_E\), and \(\Delta_t\), respectively. The left heatmap gives the validated improvements relative to the reference design, and the right heatmap gives the corresponding reference-normalized parameter values. Parameters outside \(\mathcal{P}_k\) remain at \(100\%\).}
    \label{fig:representative_candidate_profiles}
\end{figure}

\begin{figure}[ht]
    \centering
    \includegraphics[width=\linewidth]{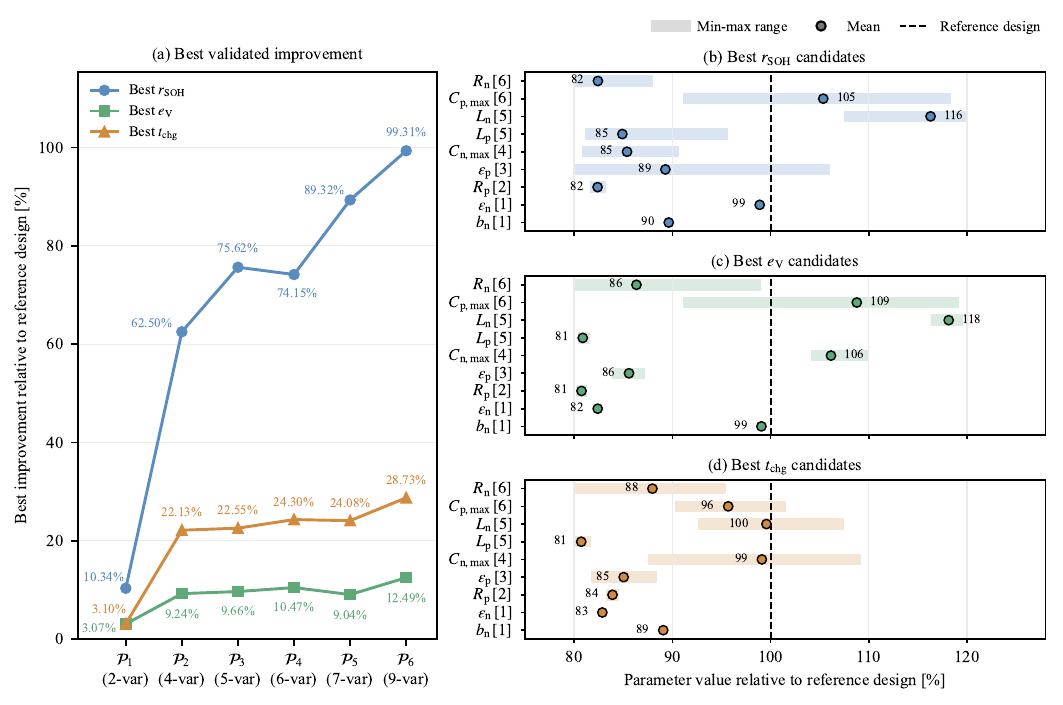}
    \caption{Best validated improvements and cross-set variation in the corresponding designs. Panel (a) reports the largest \(\Delta_{\mathrm{SOH}}\), \(\Delta_E\), and \(\Delta_t\) among the jointly improving candidates for each \(\mathcal{P}_k\). Panels (b)--(d) show the min--max ranges and means of the associated reference-normalized parameters. Bracketed numbers give the number of contributing parameter sets, and dashed lines denote the reference design.}
    \label{fig:best_improvement_parameter_stability}
\end{figure}

Surrogate-assisted Pareto optimization is performed for each reduced parameter set in Table~\ref{tab:rank_union_variable_sets} using the NSGA-II procedure described in Section~\ref{subsec:nsga2_optimization}. For every set, five independent searches are conducted with a population of 400 over 400 generations, and the first non-dominated fronts from the five runs are pooled. Candidates are submitted to the original simulation workflows only if the surrogate models predict improvement over the reference design in all three performance metrics and a non-negative SOH loss. Following this re-evaluation, a design is retained in Figs.~\ref{fig:strict_pass_improvement_distributions}--\ref{fig:best_improvement_parameter_stability} only when the simulations remain feasible and all three improvements are confirmed. The reported improvements are defined as
\begin{equation}
\begin{aligned}
\Delta_{\mathrm{SOH}}
&=
\frac{r_{\mathrm{SOH},0}-r_{\mathrm{SOH}}}
{r_{\mathrm{SOH},0}}
\times 100\%,
\\
\Delta_E
&=
\frac{e_V-e_{V,0}}
{e_{V,0}}
\times 100\%,
\\
\Delta_t
&=
\frac{t_{\mathrm{chg},0}-t_{\mathrm{chg}}}
{t_{\mathrm{chg},0}}
\times 100\%,
\end{aligned}
\label{eq:validated_relative_improvements}
\end{equation}
where the subscript \(0\) denotes the reference design. Positive values therefore indicate lower SOH loss, higher stack-level volumetric energy density, and shorter charging time, respectively. Surrogate predictions are used for screening and search. All numerical improvements discussed below are obtained from the subsequent physics-based re-evaluation.

\paragraph{(a) Jointly improving designs and parameter distributions}
Joint improvement is already possible with the two-parameter set \(\mathcal{P}_1\), but the attainable region is narrow, especially for charging time. The median improvements are \(9.40\%\) in SOH loss, \(2.67\%\) in volumetric energy density, and only \(1.39\%\) in charging time (Fig.~\ref{fig:strict_pass_improvement_distributions}g--i). Releasing the two electrode thicknesses in \(\mathcal{P}_2\) changes the balance: the median energy-density and charging-time improvements rise to \(3.92\%\) and \(13.04\%\), respectively. The corresponding medians remain near \(4\%\) and \(12\%{-}13\%\) through \(\mathcal{P}_4\), whereas the median SOH-loss reduction stays below \(9\%\). The additional particle and porous-electrode variables in \(\mathcal{P}_5\) and \(\mathcal{P}_6\) have a different effect. Median SOH-loss reduction increases to \(24.17\%\) and \(39.04\%\), and median charging-time improvement approaches \(20\%\); the median energy-density gain remains modest, at \(2.33\%\) and \(4.21\%\). Added design freedom therefore provides disproportionate leverage for degradation and charging rather than a comparable shift in all three metrics. The medians need not vary monotonically because each distribution is conditioned on a different Pareto candidate population and on passing all three simulation-based improvement tests.

The parameter windows in Fig.~\ref{fig:strict_pass_improvement_distributions}a--f are strongly structured. The median \(R_{\mathrm{n}}\) remains between \(0.80\) and \(0.90\) of its reference value for every set, and \(L_{\mathrm{p}}\) moves to \(0.80{-}0.82\) as soon as it becomes adjustable. In contrast, \(L_{\mathrm{n}}\) is consistently above the reference value, with medians of \(1.07{-}1.13\). The preferred \(c_{\mathrm{p,max}}\) is above the reference value in \(\mathcal{P}_2\)--\(\mathcal{P}_4\), but returns close to it after particle size and porosity are released. The later sets also favor smaller \(R_{\mathrm{p}}\) and lower electrode porosities. These shifts describe a coordinated rebalancing rather than independent one-parameter effects: a thinner positive electrode reduces stack volume and through-plane transport length, whereas changes in maximum concentration, negative-electrode loading, and active-material fraction compensate for the associated capacity change. Likewise, reducing \(R_{\mathrm{n}}\) shortens solid-state diffusion distances and moderates particle stress, but increases specific interfacial area. Its consistent downward shift is therefore a net result of the degradation mechanisms and operating protocols adopted here, not a universal prescription to minimize particle size. The broad ranges of \(c_{\mathrm{p,max}}\) and \(c_{\mathrm{n,max}}\) further show that several capacity-balanced combinations can produce jointly improving designs; the median profile should not be read as a unique optimum.

\paragraph{(b) Objective-specific best designs and performance trade-offs}
The representative designs in Fig.~\ref{fig:representative_candidate_profiles} expose the trade-offs hidden by the marginal distributions. Within \(\mathcal{P}_1\), the largest individual improvements are \(10.34\%\), \(3.07\%\), and \(3.10\%\) for SOH loss, energy density, and charging time, respectively. Expanding to \(\mathcal{P}_2\) produces the largest single step in performance. The degradation-best design reduces SOH loss by \(62.50\%\) while increasing energy density by \(2.67\%\) and shortening charging time by \(16.57\%\); the energy-density-best design reaches \(9.24\%\), but its accompanying improvements are only \(6.72\%\) and \(2.69\%\); and the charging-time-best design reaches \(22.13\%\) while retaining a small \(0.82\%\) energy-density gain. The best values continue to improve through the larger sets, but the preferred designs remain objective dependent.

This separation becomes especially clear in \(\mathcal{P}_6\). Its degradation-best candidate reduces the modeled 200-cycle SOH loss by \(99.31\%\), with simultaneous improvements of \(4.00\%\) in energy density and \(12.95\%\) in charging time. The energy-density-best candidate reaches \(12.49\%\), but reduces SOH loss by only \(0.62\%\) and charging time by \(5.89\%\). Conversely, the charging-time-best candidate shortens charging by \(28.73\%\) and reduces SOH loss by \(43.48\%\), while its energy-density gain is \(1.84\%\). No one of these designs replaces the Pareto set. A particularly narrow trade-off appears in \(\mathcal{P}_5\): its degradation-best and charging-time-best candidates have nearly identical parameter profiles and achieve, respectively, \((89.32\%,\,0.55\%,\,23.99\%)\) and \((88.55\%,\,0.24\%,\,24.08\%)\) in \((\Delta_{\mathrm{SOH}},\,\Delta_E,\,\Delta_t)\). In this region, degradation resistance and charging time can be improved together, but only by accepting a marginal energy-density gain.

\paragraph{Attainable performance and design trends across parameter sets}
Figure~\ref{fig:best_improvement_parameter_stability}a quantifies what is gained by increasing the optimization dimension. The transition from \(\mathcal{P}_1\) to \(\mathcal{P}_2\) is decisive: the best SOH-loss, energy-density, and charging-time improvements increase from \(10.34\%\), \(3.07\%\), and \(3.10\%\) to \(62.50\%\), \(9.24\%\), and \(22.13\%\), respectively. Thus, the four variables in \(\mathcal{P}_2\) already recover approximately \(63\%\), \(74\%\), and \(77\%\) of the respective best improvements observed in \(\mathcal{P}_6\). By \(\mathcal{P}_4\), these fractions reach approximately \(75\%\), \(84\%\), and \(85\%\). Energy density and charging time consequently exhibit clear diminishing returns beyond the four- to six-parameter sets. The degradation objective does not: its best reduction rises from \(74.15\%\) in \(\mathcal{P}_4\) to \(89.32\%\) in \(\mathcal{P}_5\) and \(99.31\%\) in \(\mathcal{P}_6\). Particle size, porosity, and negative-electrode transport provide limited additional leverage for the best energy-density design, but they are consequential when early-cycle degradation is pushed toward its minimum.

Small reversals are visible despite the overall upward trend: the best SOH-loss reduction decreases from \(75.62\%\) in \(\mathcal{P}_3\) to \(74.15\%\) in \(\mathcal{P}_4\), the best energy-density gain decreases from \(10.47\%\) in \(\mathcal{P}_4\) to \(9.04\%\) in \(\mathcal{P}_5\), and the best charging-time gain changes from \(24.30\%\) to \(24.08\%\). These are not physical penalties caused by releasing an additional parameter. Because the sets are nested, every design feasible in \(\mathcal{P}_k\) remains feasible in \(\mathcal{P}_{k+1}\) when the new variables are held at their reference values. The reversals instead reflect finite stochastic search, surrogate approximation error, and the subsequent feasibility and joint-improvement filtering. They are small relative to the principal gains, but they should not be interpreted as evidence that a lower-dimensional design space has a superior mathematical optimum.

The cross-set parameter ranges in Fig.~\ref{fig:best_improvement_parameter_stability}b--d distinguish persistent directions from objective-specific compensation. For the energy-density-best candidates, \(L_{\mathrm{p}}\) remains within \(80\%{-}82\%\) of the reference value and \(L_{\mathrm{n}}\) within \(116\%{-}120\%\), with means of \(81\%\) and \(118\%\), respectively. The degradation-best candidates consistently use a smaller \(R_{\mathrm{n}}\), spanning \(80\%{-}88\%\) with a mean of \(82\%\), while their \(L_{\mathrm{n}}\) averages \(116\%\). The charging-time-best candidates again keep \(L_{\mathrm{p}}\) within \(80\%{-}82\%\), but their \(L_{\mathrm{n}}\) remains centered near the reference value. In contrast, \(c_{\mathrm{p,max}}\) varies broadly among the degradation- and energy-density-best designs, confirming that its high Sobol index does not translate into one robust direction of change. Statistics for parameters introduced only in \(\mathcal{P}_4\)--\(\mathcal{P}_6\) are based on one to three designs and support only a directional reading, not a claim of parameter stability.

Several best candidates lie at or very close to the prescribed \(\pm20\%\) bounds, most consistently for \(L_{\mathrm{p}}\) and, in objective-dependent cases, for \(R_{\mathrm{n}}\), \(R_{\mathrm{p}}\), \(L_{\mathrm{n}}\), and electrode porosity. The optimization has therefore identified strong local directions, but not necessarily interior manufacturing optima. Extending the ranges would require new simulation data and explicit constraints on areal loading, electrode capacity ratio, mechanical integrity, electrolyte accessibility, and cell format; otherwise, independent parameter variation can generate combinations that are numerically admissible but difficult to manufacture. Finally, re-evaluation with the original electrochemical models confirms that the surrogate-derived gains survive the high-fidelity simulation workflows. It does not remove model-form uncertainty or replace experimental validation. Within those limits, \(\mathcal{P}_2\) captures most of the observed energy-density and charging-time gains, whereas the strongest reduction in modeled early-cycle SOH loss appears only after the particle, porosity, and transport variables are released in \(\mathcal{P}_5\) and \(\mathcal{P}_6\).

\section{Conclusion}
Further advances in electric mobility increasingly depend on battery cells that deliver more usable energy and faster charging without compromising durability. This study has developed a timescale-aware surrogate-assisted framework for jointly optimizing stack-level volumetric energy density, fast-charging capability, and SOH loss. By combining objective-specific surrogate models, cross-objective rank-union variable sets derived from Sobol analysis, Pareto searches over nested variable sets, and physics-based re-evaluation, the framework accommodates the markedly different evaluation costs and physical timescales of the three performance metrics. This objective-wise formulation makes it computationally tractable to retain multi-cycle SOH loss as an explicit design objective rather than replacing degradation with a BoL proxy.

Parameter importance varies strongly across the three objectives, and single-metric screening is therefore inadequate for joint cell design. The surrogate-model comparison revealed a similar objective dependence: the preferred model structure differed among the three responses, and SOH loss was more difficult to approximate than either BoL metric. Physics-based re-evaluation verified jointly improving candidates in every nested variable set. Among these candidates, the objective-wise best solutions, attained by different designs, reduced SOH loss by 99.31\%, increased stack-level volumetric energy density by 12.49\%, and shortened charging time by 28.73\%. These findings show that degradation, energy density, and charging performance can be treated as coequal design objectives even when their evaluation requires markedly different simulation horizons. The nested Pareto searches further revealed earlier diminishing returns in energy density and charging time, whereas achieving the strongest SOH-loss reduction required additional particle-size, porosity, and transport degrees of freedom. Future work can extend the framework by incorporating manufacturability constraints, additional degradation mechanisms, and cell-level experimental validation, thereby advancing the identified design directions toward practical cell development.

\section*{Acknowledgement}
This work was financially supported by the Swedish Research Council under Grant No.~2023-04314. Open access funding provided by Chalmers University of Technology


\bibliography{Reference}

\end{document}